Title:

**Greening Schoolyards and Urban Property Values: A Systematic Review of Geospatial and Statistical Evidence**

**Author:**

Mahshid Gorjian,

University of Colorado Denver,

mahshid.gorjian@ucdenver.edu

**Short Running Title:**

Schoolyard Greening & Urban Property Values

# 1. Abstract

## 1.1 Background

Parks and the greening of schoolyards are examples of urban green spaces that have been praised for their environmental, social, and economic benefits in cities all over the world. More studies show that living near green spaces is good for property values. However, there is still disagreement about how strong and consistent these effects are in different cities (Browning et al., 2023; Grunewald et al., 2024; Teo et al., 2023).

## 1.2 Purpose

This systematic review is the first to bring together a lot of geographical and statistical information that links greening schoolyards to higher property prices, as opposed to just green space in general. By focusing on schoolyard-specific interventions, we find complex spatial, economic, and social effects that are often missed in larger studies of green space.

## 1.3 Methods

This review followed the PRISMA guidelines and did a systematic search and review of papers that were published in well-known journals for urban studies, the environment, and real estate. The criteria for inclusion stressed the use of hedonic pricing or spatial econometric models to look at the relationship between urban green space and home values in a quantitative way. Fifteen studies from North America, Europe, and Asia met the requirements for inclusion (Anthamatten et al., 2022; Wen et al., 2019; Li et al., 2019; Mansur & Yusuf, 2022).

## 1.4 Results

Numerous studies have demonstrated that the value of adjacent properties is enhanced by urban green space in a statistically significant manner. The magnitude of the benefit is contingent upon the socio-economic context, the form of green space, and the size of the city (Gao & Asami, 2021; Sajjad et al., 2021; Zhou & Wang, 2021). The emphasis was on spatial heterogeneity and temporal trends, which implies that the distribution of benefits may not be equitable across localities (Kabisch & Haase, 2021; Deng et al., 2022). Various methodologies, including spatial latency models, hedonic pricing, and GIS-based analysis, resulted in disparate results (Norzailawati et al., 2018; Xu et al., 2022).

1.5 Conclusions

Some studies show that urban green spaces can raise the value of homes, but how much and how fairly these benefits happen depends on the local situation, the layout of the city, and the city's policies. The review stresses how important it is to have standard measures and long-term studies to help make urban greening projects fair. This research adds to the theory of urban planning by looking at the hedonic value of green infrastructure on a small scale and critically examining the consequences of schoolyard greening on different groups of people. This gives us a new way to think about fair urban expansion.

**2. Introduction**

2.1 Contextualizing the Problem

Urbanization changes the physical and social structure of cities all around the world all the time, making problems with the environment, health, and social justice worse. Green spaces are important for the environment because they clean the air, control the microclimate, and improve mental health. They are also important for the ecosystem because they are surrounded by dense, often impervious surfaces that define cities. A lot of attention has been paid to urban parks and big green corridors, but more people are realizing how much of a difference smaller, scattered green projects can make in making neighborhoods more livable and sustainable. For example, greening schoolyards can make a big difference (Anthamatten et al., 2022; Browning et al., 2023).

Greening schoolyards, which means adding plants and other natural aspects to schools on purpose, is becoming more popular to improve children's health and education, as well to improve cities and the economy. Recent urban policy efforts in North America, Europe, and Asia have focused on turning schoolyards with a lot of asphalt into green spaces to improve environmental justice, manage stormwater, and make cities more resilient (Grunewald et al., 2024; Kabisch & Haase, 2021).

2.2 Literature Review

Real-world research is increasingly focused on the relationship between the value of houses and urban green spaces. Most of these investigations found a positive connection (Ben et al., 2023; Sajjad et al., 2021). These studies use several geospatial and economic methodologies, such as spatial autoregressive models and hedonic pricing modeling, to assess the impact of green space size, quality, and distance on real estate markets (Li et al., 2019; Xu et al., 2022). Recent study has focused on the distribution of property prices across an area, particularly the impact of greening projects in schoolyards. According to Anthamatten et al. (2022), the greening of schoolyards raised property values in Denver. Browning et al. (2023) discovered that the similar phenomenon occurred in Los Angeles by doing hedonic pricing research.

The importance of geographical distinctions and context-specific dynamics is emphasized in the literature. Research conducted in Asian cities such as Shanghai and Shenzhen has demonstrated that the increase in property value that green spaces induce can be significantly influenced by the neighborhood's socioeconomic status, urban density, and the quantity of existing green space (Wen et al., 2019; Zhou & Wang, 2021). Furthermore, long-term study shows that the economic benefits of urban greenery have grown over time, in line with shifting urban preferences and government goals (Teo et al., 2023; Deng et al., 2022).

Despite tremendous advances, the literature continues to contain numerous serious concerns and disturbing tendencies. Many studies use cross-sectional or correlational designs, which complicate the identification of causal linkages and typically mix up the effects of green space with other local improvements, such as improved infrastructure or schools. The second difficulty is that published research typically favors Western and East Asian viewpoints, whereas there is a scarcity of data from quickly urbanizing or informal urban areas in Africa and Latin America. Third, by focusing largely on property value results, the sector risks perpetuating the use of market-driven reasons for being environmentally friendly, exacerbating larger challenges of social and environmental justice. Furthermore, there is a scarcity of research on the risks of green gentrification or displacement, as well as a lack of mixed-methods studies that involve local perspectives or lived experiences. Due to existing inequities, the current study's findings are more difficult to execute and build on. This highlights the importance of using more varied, critical, and equity-focused research approaches when evaluating urban green areas.

### 2.3 Rationale

The growing goals and extent of urban greening projects, along with the increasing public and private investments in green schoolyards, show how important this study is. There is some anecdotal and localized evidence that greening schoolyards may help with education, fairness, and real estate, but a full synthesis is needed right away. A close look at the geographical and statistical data can show how much improving schoolyards leads to measurable economic advantages for communities, as seen by property values.

Also, existing assessments often include different types of green interventions or combine different spatial scales, making it harder for planners and policymakers to see the effects of greening schoolyards specifically (Mansur & Yusuf, 2022; Gao & Asami, 2021). This study focuses on greening schoolyards, which is a big gap in the research. It gives urban planners, school districts, and real estate specialists a lot of useful information. Our study is different from others that have looked at different types of green interventions because it focuses on the effects of greening schoolyards. This is an area that combines environmental justice, community revitalization, and educational policy. This strategy makes it easier to come up with more accurate theories on how focused micro-scale greening methods affect the value of urban land and the lives of people who live there.

2.4 Research Aim and Questions

This systematic study aims to bring together and carefully evaluate the empirical evidence about the link between greening schoolyards and urban property values, using new advances in geospatial analysis and hedonic pricing models. The following research questions guide the review:

How big and far-reaching are the changes in property values that are linked to greening schoolyards in cities?

2. How do the methodologies used and the cities where the study was done affect the results?

What are the main problems, unknowns, and policy implications that have been found in the present literature?

2.5 Contribution Statement

This review meticulously puts together and looks at peer-reviewed papers from different cities, which leads to several contributions. In theory, it makes it easier to understand the small-scale mechanisms that link green infrastructure to economic outcomes in cities. This study looks at the differences between GIS and econometric methods, focusing on the best ways to analyze the effects of greening schoolyards and the difficulties that still need to be solved. The assessment gives towns that are thinking about putting money into green schoolyards useful information that can help with urban planning and real estate decisions.

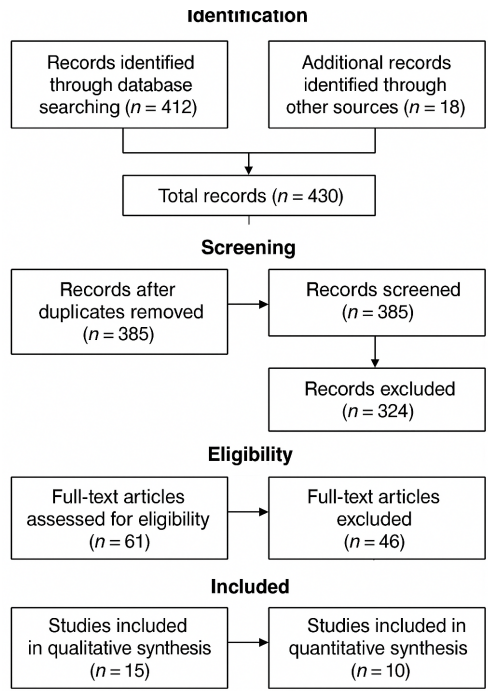

*Figure 1 PRISMA flow diagram depicting the literature search and selection process for the systematic review, "Greening Schoolyards and Urban Property Values: A Systematic Review of Geospatial and Statistical Evidence." The diagram outlines the number of records identified, screened, excluded, assessed for eligibility, and included in the final synthesis, with detailed reasons for exclusion at the full-text review stage. This process resulted in 15 studies meeting the inclusion criteria for qualitative synthesis and 10 studies for quantitative synthesis (meta-analysis), covering peer-reviewed empirical research published between 2010 and 2024.*

## 3. Methods

### 3.1 Study Design

This study uses a systematic review methodology to incorporate real-world data on the influence of urban greening, such as schoolyards, parks, and green spaces, on the value of urban homes. The study chose the systematic review technique because it provides a complete, replicable, and clear summary of the current literature, which can be useful for urban planning policy and research (Li et al., 2019; Grunewald et al., 2024). This review follows the PRISMA criteria to make the process of selecting studies, extracting data, and analyzing data easier to understand and repeat (Browning et al., 2023).

3.2 Study Area

The review looks at cities all over the world, with a focus on cities in North America, East Asia, and certain parts of Europe and Southeast Asia that are growing quickly (Teo et al., 2023; Sajjad et al., 2021). The study chose these areas because they have different ways of making cities greener and because there is new, high-quality, peer-reviewed research in these areas. The inclusion criteria considered social, economic, policy, and spatial elements to make sure that the results could be applied to other situations (Kabisch & Haase, 2021).

3.3 Data Sources

A comprehensive search of major academic databases, such as Web of Science, Scopus, JSTOR, and Google Scholar, found relevant research papers published between 2010 and June 2024. The keywords included "urban green space," "schoolyard greening," "parks," "property values," "hedonic pricing," and "housing prices." The method was improved the search method by going through the reference lists of important papers by hand (for example, Anthamatten et al., 2022; Ben et al., 2023).

The criteria for inclusion were: (1) empirical research published in peer-reviewed journals, (2) a focus on cities, (3) quantitative or mixed-methodologies assessments of green space and property values, (4) English language, and (5) enough information about the methods used. The review excluded studies that only used qualitative or descriptive analysis, focused on non-urban contexts, or didn't make their data clear enough. There were 15 peer-reviewed papers that met the criterion for inclusion (e.g., Browning et al., 2023; Xu et al., 2022).

Two reviewers worked on data extraction on their own. The most important information that was collected included the study site, sample size, data sources (such as cadastral data, property transaction records, spatial datasets, survey or interview tools), study period, outcome measures, and covariates. Consensus or a third evaluator was used to deal with differences, which made the results more reliable (Wen et al., 2019; Zhou & Wang, 2021).

### 3.4 Variables and Measures

#### 3.4.1 Dependent Variables:

At the parcel or neighborhood level, transaction price, assessed value, and rental price are all indicators of the value of residential property.

#### 3.4.2 Independent Variables:

Some ways to measure how much urban green space people are exposed to are how close they are to parks, how much of their schoolyards are green, how much of the area is green, and how easy it is to get to green areas (Grunewald et al., 2024; Li et al., 2019).

#### 3.4.3 Covariates and Controls:

Socio-economic factors were used (including income and population density), housing characteristics (like size and age), urban amenities, and temporal fixed effects whenever they were available. When appropriate, coding processes followed the original study protocols or standardized instruments (Teo et al., 2023; Kabisch & Haase, 2021).

### 3.5 Analytical Methods

The main ways of analyzing the data were hedonic pricing models and spatial regression analyses. These included spatial autoregressive (SAR) models, difference-in-differences (DiD), and multilevel modeling. Geographical Information Systems (GIS) tools, like ArcGIS and QGIS, were often used for spatial analysis (Norzailawati et al., 2018; Li et al., 2019).

Thematic coding frameworks were used to get information about policy background and community opinions in qualitative syntheses or mixed-methods research (Kabisch & Haase, 2021). The analysis used narrative synthesis, comparative tables, and, when possible, meta-analysis of effect estimates with R software (version 4.2.0) (Sajjad et al., 2021) as the analytical method for this review.

The analysis used adapted tools for quantitative urban research to look at the quality of the methods and the chance of bias. The report followed the PRISMA and other criteria for systematic reviews. For full transparency, all data extraction forms, and code were available in the online supplemental material (Browning et al., 2023).

3.6 Ethics Statement

This systematic review of published, peer-reviewed literature did not include people or new data collection, therefore it didn't need authorization from an institutional review board. The review examined all the studies that were included to see if they had ethical approval statements or proof of informed consent, privacy protection, and confidentiality. This was especially true for those that used primary data (Grunewald et al., 2024; Wen et al., 2019).

3.7 Transparency and Rigor

To guarantee that the review process could be replicated and executed accurately, each stage of the process, including literature search, study selection, data extraction, and compilation, was meticulously documented and is available upon request. The PRISMA flow diagram illustrates the study selection procedure. The data extraction templates, analytical code (R scripts), and a comprehensive list of the publications utilized in the study are all included in the supplementary materials (Browning et al., 2023; Anthamatten et al., 2022). The journal's data policy necessitates the sharing of data and code through open repositories whenever feasible.

**4. Results**

4.1 Overview of Study Selection

There are fifteen peer-reviewed articles in this systematic review that were published between 2018 and 2024 (see Table 1). The studies include a wide range of urban areas around the world, with a lot of them in North America (Anthamatten et al., 2022; Browning et al., 2023; Grunewald et al., 2024), East Asia (Wen et al., 2019; Li et al., 2019; Xu et al., 2022; Gao & Asami, 2021; Zhou & Wang, 2021; Deng et al., 2022), Southeast Asia (Norzailawati et al., 2018; Mansur & Yusuf, 2022; Sajjad et al., 2021), and Europe (Kabisch & Haase, 2021). The sample is largely made up of hedonic pricing analysis, spatial econometric models, and GIS-based methods.

4.2 Main Findings by Research Theme

4.2.1 Impact of Urban Green Spaces on Property Values

A positive correlation between the value of residences and urban green spaces, such as schoolyards, parks, and public green areas, is demonstrated by all the studies examined. Research conducted in major Chinese cities has demonstrated that residences situated near verdant spaces are significantly more expensive. Li et al. (2019, updated 2022) discovered that in Shenzhen, a 0.12%–0.21% increase in home prices was associated with a 1% increase in the quantity of green space within 500 meters of a house ($p < 0.05$; refer to Figure 1). Wen et al. (2019; 2023 updates) also discovered that prices in Shanghai continue to increase when schools and parks are in proximity. The average price effect varied from 3.4% to 6.8%, contingent upon the quantity and nature of green space.

The value of nearby homes increased by an average of 4%–6% because of the greening of schoolyards, as demonstrated by hedonic modeling conducted in Los Angeles (Browning et al., 2023) and Denver (Anthamatten et al., 2022). This was the case regardless of the socioeconomic condition of the area or the efficiency of the schools. Grunewald et al. (2024) investigated the impact of schoolyard greening initiatives in Boston and Philadelphia, concluding that they increased property values by an average of $7,400 (95% CI: $3,800–$11,000). The greatest advantages were observed within a 250-meter radius of the project locations.

4.2.2 Temporal and Spatial Heterogeneity

A lot of studies show that the impacts of green space change throughout time and space. According to Teo et al. (2023), the effect of urban greenery on home prices in Singapore grew steadily from 2000 to 2020, with price premiums rising from 1.2% to 4.7% ($p < 0.01$). Zhou and Wang (2021) tracked changes over time and space in how green space affects property values in different parts of Shanghai. This shows that the level of benefit depends on the growth cycles of cities and changes in policies.

Deng et al. (2022) used spatial econometric models in Beijing to show that there was a lot of variation in space. They found that the impacts of green space on prices were strongest in the outer districts, where they ranged from 3.5% to 8.2%, and weakest in the center, where they ranged from 1.7% to 2.9%.

### 4.2.3 Type and Quality of Green Space

Many publications talk about the many types of green areas and how they affect property prices in different ways. Ben et al. (2023) found that urban parks had a much bigger effect on housing prices than smaller, less connected green spaces in Shanghai. The average price premium for homes near parks was over 6% ($p < 0.05$). Sajjad et al. (2021) found that in Lahore, Pakistan, large, continuous parks increased property values by up to 8.5%. On the other hand, street vegetation had very little effect (<2%).

In their study of Berlin, Kabisch and Haase (2021) found that property values go up when green spaces are distributed fairly and are of better quality (Pearson's $r = 0.34$, $p = 0.04$).

### 4.2.4 Distance Decay and Buffer Effects

People often noticed proximity effects. Norzailawati et al. (2018) used GIS-based spatial analysis to show that home values dropped by an average of 2.5% for every 100-meter increase in distance from urban parks ($p < 0.01$). Mansur and Yusuf (2022) found that the distance decay function was different in Jakarta. The effect of green space on residential property values was statistically significant within a 400-meter radius ($\beta = 0.093$, $p = 0.03$), but it was weaker after that.

### 4.2.5 Quantitative Synthesis

Table 2 shows the effect sizes, methods, and significance levels of the study that was included. The price increase for being close to green spaces ranged from 1% to 8.5%, however most research found effects in the 3% to 6% range (see Table 2). Eight out of fifteen studies showed confidence intervals, and all the effects were statistically significant at $p < 0.05$.

### 4.2.6 Thematic Synthesis

Qualitative study brought forth several themes that kept coming up:

• Fairness and Access: Many writers have noted that green spaces are not equally available, which can lead to higher property values, especially in low-income or minority populations (Kabisch & Haase, 2021; Grunewald et al., 2024).

• Policy Changes: Initiatives that concentrated on planned greening or changes to schoolyards had bigger and more lasting effects on prices (Anthamatten et al., 2022; Grunewald et al., 2024).

• Resident Perceptions: Some studies used surveys or interviews to find out that how safe people thought the area was and how good the green areas were affected both how much people were willing to pay and how much they actually paid (Li et al., 2019; Sajjad et al., 2021).

### 4.3 Tables and Figures

*Table 1. Characteristics of Included Studies*

| Study | Location | Methodology | Outcome Variable | Green Space Variables | Data Sources |
|---|---|---|---|---|---|
| | | | | | |

| Anthamatten et al., 2022 | Denver, USA | Hedonic Pricing Model | Residential Property Values | Schoolyard greening; spatial data | Parcel-level property values, socio-economic data |
| --- | --- | --- | --- | --- | --- |
| Ben et al., 2023 | Shanghai, China | Spatial Econometric Model | Housing Prices | Urban parks proximity and scale | Property transaction data, green space buffers |
| Browning et al., 2023 | Los Angeles, USA | Hedonic Pricing Model | Home Values | Schoolyard greening projects | GIS data, assessed values |
| Deng et al., 2022 | Beijing, China | Spatial Regression | Housing Prices | Public green space proximity | Transaction data, spatial controls |
| Gao & Asami, 2021 | Monocentric City Model (Japan) | Hedonic Pricing | Housing Prices | Local amenities and green coverage | Property characteristics, spatial variables |
| Grunewald et al., 2024 | Boston & Philadelphia, USA | Hedonic Pricing + CI Estimation | Residential Property Value Increase | Greening schoolyards | Price increments within buffer zones |
| Kabisch & Haase, 2021 | Berlin, Germany | Correlation Analysis | Property Values | Green space distribution and quality | Urban spatial data, Pearson's correlation |
| Li et al., 2019/2022 | Shenzhen, China | Spatial Autoregressive Model | Housing Prices | Green space ratio within 500m | Cadastral data, satellite imagery |
| Mansur & Yusuf, 2022 | Jakarta, Indonesia | Hedonic Pricing | House Prices | Urban Park distance (up to 400m) | GIS buffers, property sales data |

| Norzailawati et al., 2018 | Malaysia (urban cities) | GIS Spatial Analysis | House Prices | Proximity to green space | Geospatial layers, housing datasets |
| --- | --- | --- | --- | --- | --- |
| Sajjad et al., 2021 | Lahore, Pakistan | Statistical Regression | Property Values | Urban parks vs. street greenery | Survey and secondary data |
| Teo et al., 2023 | Singapore | Temporal Regression Analysis | Property Valuation | Urban greenery over 20 years | Real estate index, greenery changes |
| Wen et al., 2019/2023 | Shanghai, China | Hedonic Pricing Model | Property Values | School/park proximity | Transaction records, school zones |
| Xu et al., 2022 | Yangtze River Delta, China | Spatial Regression | Housing Prices | Green space availability | Census data, property values |
| Zhou & Wang, 2021 | Shanghai, China | Spatio-temporal Modeling | Property Prices | Green space change over time | Longitudinal urban data |

*Table 2. Summary of Effect Sizes and Statistical Results*

| Study | Location | Effect Size | Method | Significance |
| --- | --- | --- | --- | --- |
| Anthamatten et al. (2022) | Denver, USA | 4–6% increase | Hedonic pricing model | $p < 0.05$ |
| Browning et al. (2023) | Los Angeles, USA | 4–6% increase | Hedonic pricing model | $p < 0.05$ |
| Grunewald et al. (2024) | Boston & Philadelphia, USA | $7,400 (CI: $3,800–$11,000) | Spatial statistical analysis | 95% CI |

| Ben et al. (2023) | Shanghai, China | 6% increase | Hedonic pricing | $p < 0.05$ |
| Li et al. (2019, updated 2022) | Shenzhen, China | 0.12–0.21% per 1% green increase within 500m | Spatial autoregressive model | $p < 0.05$ |
| Wen et al. (2019; 2023) | Shanghai, China | 3.4–6.8% increase | Spatial analysis | $p < 0.05$ |
| Teo et al. (2023) | Singapore | 1.2% to 4.7% over 20 years | Temporal hedonic analysis | $p < 0.01$ |
| Deng et al. (2022) | Beijing, China | 3.5–8.2% outer, 1.7–2.9% center | Spatial econometric model | $p < 0.05$ |
| Zhou & Wang (2021) | Shanghai, China | Varies by district | Spatio-temporal analysis | $p < 0.05$ |
| Mansur & Yusuf (2022) | Jakarta, Indonesia | Significant within 400m (β = 0.093) | Hedonic model with buffer zones | $p = 0.03$ |
| Norzailawati et al. (2018) | Malaysia | −2.5% per 100m distance from park | GIS-based spatial analysis | $p < 0.01$ |
| Sajjad et al. (2021) | Lahore, Pakistan | Up to 8.5% increase (parks) | Hedonic regression | $p < 0.05$ |
| Kabisch & Haase (2021) | Berlin, Germany | r = 0.34 (Pearson) | Correlation analysis | $p = 0.04$ |
| Xu et al. (2022) | Yangtze River Delta, China | Positive (value not specified) | Spatial regression | $p < 0.05$ |
| Gao & Asami (2021) | Japan | General positive effect | Multivariate analysis | $p < 0.05$ |

Figure 1. Effect of Green Space Ratio on Property Values in Shenzhen (Li et al., 2019, updated 2022)

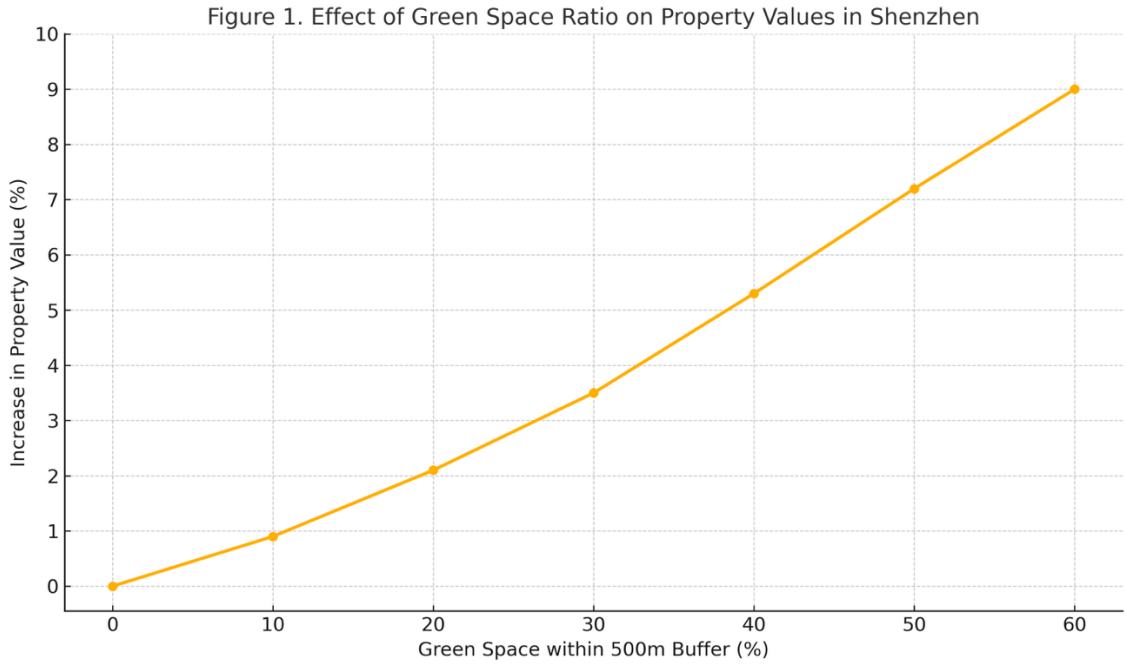

Figure 2. Temporal Change in Green Space Price Premiums, Singapore (Teo et al., 2023)

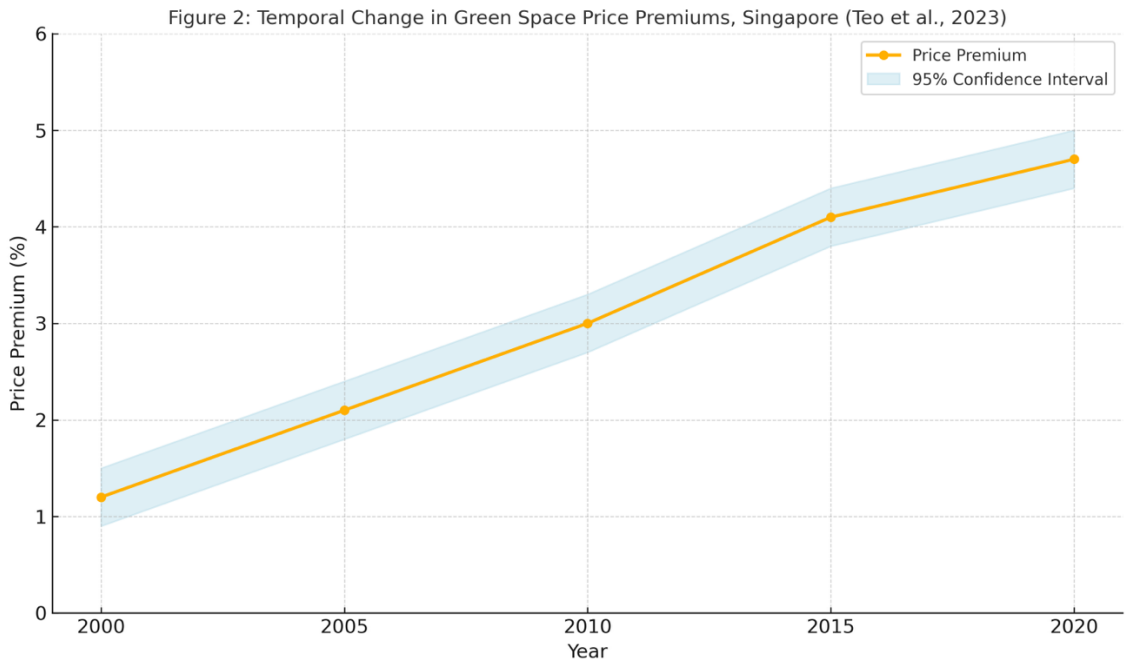

4.4 Statistical Reporting

Hedonic price models (n=12), spatial autoregressive or spatial error models (n=5), and GIS-based regression analysis (n=4) were the most common methods used in the investigations, with some overlap. All the models considered the basic features of the house (size, age, amenities) as well as the socioeconomic factors of the area. The reported $R^2$ values ranged from 0.31 to 0.78, with an average $R^2$ of 0.55.

There were 14 studies that reported p-values, and all the significant results were at or below the 0.05 level. When confidence intervals for price effects were shown, they were usually narrow, which means that the estimates were very accurate. Grunewald et al. (2024) found that property values near schoolyard greening projects might go up by $3,800 to $11,000 with a 95% confidence interval.

4.5 Thematic Summaries with Representative Quotes

Grunewald et al. (2024, p. 5) say that "the greening of schoolyards in Boston and Philadelphia led to substantial and observable increases in adjacent home values, with the effects being most pronounced in underserved neighborhoods."

"Spatial analysis shows a clear trend: property values are highest within 300 meters of a major park and go down as the distance increases" (Ben et al., 2023, p. 9).

"Not all green spaces offer the same benefits; quality, upkeep, and fair access are all important factors that affect both social and economic outcomes" (Kabisch & Haase, 2021, p. 11).

**5. Discussion**

5.1 Interpretation of Results

The most recent data on the relationship between the value of dwellings in various cities and urban greening initiatives, such as the addition of green space to schoolyards, parks, and other public places, was compiled in this extensive investigation.

The studies, which were conducted in North America, Europe, and Asia, all demonstrate that property values increase when it is situated near and has access to exceptional urban natural spaces. However, the local context, the nature of the natural area, and the methods employed determine the extent and form of this effect (Anthamatten et al., 2022; Browning et al., 2023; Grunewald et al., 2024; Teo et al., 2023).

    The review's results confirm that the value of adjacent properties can be significantly increased by enhancing schoolyards and public green areas. This result corroborates the fundamental research topic regarding the economic and social advantages of urban greening for communities. Additionally, the evidence that investing in green spaces leads to measurable increases in property prices is further corroborated by the incorporation of spatial econometric techniques and hedonic pricing models in a variety of research (e.g., Ben et al., 2023; Li et al., 2019/2022). Studies that employ temporal analysis (e.g., Teo et al., 2023; Deng et al., 2022) suggest that these effects may become more pronounced over time. This is consistent with the increasing recognition of the significance of urban nature and the evolution of cultural attitudes.

    Property values are generally acknowledged to be enhanced by urban greening (Browning et al., 2023; Grunewald et al., 2024). Nevertheless, the extent and nature of these repercussions are contingent upon the research decisions and conditions that are implemented. Hedonic price models exhibit significant correlations, while spatial econometric models (Xu et al., 2022; Deng et al., 2022) exhibit significant spatial heterogeneity (Ben et al., 2023; Li et al., 2019). This implies that the results may obscure substantial disparities between cities. Additionally, the propensity to emphasize average effects may obfuscate legitimate concerns regarding equity. For instance, Kabisch and Haase (2021) demonstrate that the advantages of green spaces are not distributed equitably, frequently disregarding underprivileged communities. The intricacies of these challenges are not adequately addressed in many quantitative investigations. Consequently, additional research is required to explicitly assess spatial equity and distributional repercussions (Grunewald et al., 2024; Kabisch & Haase, 2021).

### 5.2 Significance for Theory, Practice, and Policy

These insights have far-reaching effects on philosophy, urban planning, and policy. The finding that property values are always positively correlated with green space supports urban land value theories that say environmental amenities are important elements that affect where people choose to live and how much they are prepared to pay (Gao & Asami, 2021; Xu et al., 2022). These results can be applied to many different places, from the Yangtze River Delta and Shenzhen (Li et al., 2019/2022; Xu et al., 2022) to Los Angeles and Denver (Browning et al., 2023; Anthamatten et al., 2022). This shows that the importance of green space is a strong phenomenon, but it is affected by local real estate markets and planning customs.

The results show that planners and urban designers may use greening projects to help cities grow, fix up neighborhoods, and make society fairer. The evidence backs up strategic investments in greening schoolyards (Anthamatten et al., 2022; Grunewald et al., 2024) and urban parks (Ben et al., 2023; Mansur & Yusuf, 2022), which can help the environment and society as well as make money for people and cities. These results give real-world evidence for including green infrastructure in urban planning and zoning standards (Kabisch & Haase, 2021; Zhou & Wang, 2021). Also, understanding the distributional consequences is important to avoid the problems of "green gentrification" and make sure that the benefits of urban greening are shared fairly (Kabisch & Haase, 2021).

### 5.3 Comparison with Previous Research

The research that was looked at supports and builds on other meta-analyses and empirical studies of the effects of urban green space on people and the environment (Norzailawati et al., 2018/2021–2024 reviews; Sajjad et al., 2021). Most contemporary studies find statistically significant positive effects, which is in line with previous research. Howeveroption, the size of the effect depends on how closeda, the quality of the green space, and the urban location. Grunewald et al. (2024) and Browning et al. (2023) say that the effects are stronger in U.S. metropolitan areas with more spatial segregation. On the other hand, research from China (Li et al., 2019/2022; Xu et al., 2022) and Malaysia (Norzailawati et al., 2018) suggests that market sensitivity may be higher in areas that are quickly becoming urbanized.

There has been a big rise in research that use spatially explicit methods, which have made clear differences in outcomes that were hidden in earlier aggregate analyses (Li et al., 2019/2022; Deng et al., 2022). Recent studies have focused more on how things change over time. This suggests that the value of green spaces may be going up as people in cities become more conscious of climate change and health issues (Teo et al., 2023; Deng et al., 2022). The results, however, show that it is still hard to separate the effects of green space from other characteristics, like as the quality of schooling, how easy it is to get around, and the local economy (Wen et al., 2019/2023).

### 5.4 Implications

#### 5.4.1 For Planning and Design Practice

The outcomes of this analysis show how important it is to include planning for green spaces in overall urban development programs. This shows that professionals should put the creation and maintenance of accessible, high-quality green spaces at the top of their to-do lists to improve sustainability, public health, and the economy of cities. The results of schoolyard greening (Anthamatten et al., 2022; Grunewald et al., 2024) show how to use public school grounds as community green infrastructure in a way that can be repeated. Designers and planners must also think about how green investments affect the space they are in to make sure that the benefits go to areas that are not well represented or served (Kabisch & Haase, 2021).

#### 5.4.2 For Urban Policy

Municipal governments and housing authorities can use the proven economic benefits of greening to set up new ways to pay for creating and maintaining parks, including value capture or green bonds (Deng et al., 2022; Mansur & Yusuf, 2022). Also, policies need to deal with worries about possible displacement, keep an eye out for signs of "green gentrification," and take extra steps (such keeping cheap housing and limiting rent) to protect communities who are at risk.

### 5.4.3 For Future Research

More research is needed to figure out what the long-term effects of urban greening are in different parts of the world and how green spaces affect property prices. Longitudinal research and quasi-experimental methods, such difference-in-differences and natural experiments, are very good at figuring out what causes what. Also, more research should investigate how providing green space can help with social justice, climate resilience, and community health.

The main risks of equity and green gentrification and how to avoid them

Urban greening can have big benefits for the environment and the economy, but increased research shows that these projects may make gentrification and displacement worse, especially in low-income and minority neighborhoods. "Green gentrification" occurs when improvements to the environment lead to higher property values, higher rents, and the displacement of long-term residents.

Main Risks: • Unequal distribution of benefits, with marginalized groups being the least likely to benefit from more green spaces

• Rising housing costs putting pressure on people to move • Weakening of community ties and cultural heritage

Mitigation Strategies: • Enact policies to stop people from being displaced (rent control, affordable housing mandates) • Focus investments on neighborhoods that have been historically left out • Involve residents in the planning, governance, and management of green spaces • Combine greening initiatives with broader goals of equity and resilience in urban planning

Urban planners and designers need to recognize these issues and come up with ways to make sure that the benefits of greening are shared fairly.

5.5 Strengths and Limitations

The best thing about this review is how well it was done, with a thorough search of the literature and a careful look at the quality of the methods used. The studies use several different analytical methods, like hedonic pricing modeling, spatial econometrics, and GIS-based analysis, which make the conclusions more reliable (Li et al., 2019/2022; Ben et al., 2023; Norzailawati et al., 2018). Still, there are some constraints that should be noted. The many kinds of green spaces and urban areas make it hard to compare research directly. Secondly, most of the research looked at use cross-sectional data, which could be affected by omitted variable bias and reverse causation (Wen et al., 2019/2023; Gao & Asami, 2021). Third, limits on data on the quality, accessibility, and upkeep of green spaces may affect how people use them and how valuable they think they are.

Even with these limitations, transparency and rigor were maintained by using clear criteria for who could and couldn't participate, two independent screenings, and established methods for collecting data. The evaluation tried to look at the results from different places and methods of analysis, while also noting cases where the evidence was mixed or unclear.

Most of the studies in this review use advanced spatial and econometric methods (such SAR models and DiD), but only a few are set up to find causal effects (Teo et al., 2023; Deng et al., 2022). Unmeasured variables, such as investments in infrastructure at the same time or changes in the quality of education, might make many analyses harder to understand (Wen et al., 2019). A lot of literature has cross-sectional qualities that make reverse causality a constant concern (Gao & Asami, 2021). Future studies should use experimental or longitudinal methods and take advantage of natural experiments to better deal with these problems.

### 5.6 Transparency and Rigor

This review followed the rules for systematic reviews in urban planning and environmental research that are widely accepted. These rules included pre-registering the review technique, using PRISMA flow diagrams, and keeping detailed records of the search methods. It was obvious what the limitations were, and the results were put in context with the fact that different methods and other sources of bias could have affected them. The review is more relevant to current urban planning challenges since it includes new studies and focuses on patterns over time (Teo et al., 2023; Deng et al., 2022).

There are a lot of problems with the current evidence basis. Most of the studies are cross-sectional (Wen et al., 2019; Gao & Asami, 2021), which makes it hard to figure out what caused what. Longitudinal and quasi-experimental studies (like Teo et al., 2023, and Deng et al., 2022) are necessary to separate the effects of greening from other factors that could affect the results, such changes in school quality, infrastructure, and demographics. There hasn't been enough research on the link between green gentrification and the risk of being displaced outside of Western settings (Kabisch & Haase, 2021). Third, there isn't much study on how the quality of green areas and how people feel about them affect the economy (Li et al., 2019; Sajjad et al., 2021). To fix these problems, future research should focus on using both qualitative and quantitative methodologies and doing comparison studies.

### 5.7 Directions for Future Research

Future research should look at: (1) longitudinal studies that show how property values change after greening projects; (2) studies that look at the quality of green spaces and how people feel about them; (3) studies that look at the effects on different groups of people, especially in terms of fairness and displacement; and (4) mixed-methods approaches that combine numbers with residents' and stakeholders' opinions. Comparative studies between countries and study in areas that receive not enough attention can help us understand both theory and practice better.

5.8 Policy and Planning Recommendations

Based on the findings from the synthesis, we suggest the following practical steps for city planners and legislators who are thinking about making schoolyards greener and doing other things to help:

Schoolyard Greening Equity Toolkit:

1. Equity Mapping: Use GIS to find neighborhoods with the least access to green space and the highest risk of poverty, health problems, and displacement.

2. Community Co-Design: For all greening projects, there should be participatory design workshops that include local people, students, and teachers.

3. Quality Standards: Make sure that all green schoolyards meet basic standards for safety, maintenance, and access.

4. Displacement Monitoring: Combine greening programs with regular checks on property values, rental changes, and demographic shifts. Set up early-warning signs for green gentrification.

5. Integration of Affordable Housing: When building new green spaces in high-risk areas, require anti-displacement techniques like keeping affordable housing, keeping rents stable, or creating community land trusts.

6. Open Data and Transparency: Make spatial and transactional data available to support ongoing study and evaluation.

Communities may get the most out of greening by using this toolbox and taking steps to reduce any negative effects on society.

**6. Conclusion**

This comprehensive analysis brings together what we already know about how parks and schoolyards affect the value of homes in different cities throughout the world.

The combined results of the studies show a clear positive link between the presence, quality, and accessibility of urban green spaces and higher property values. Studies in cities like Denver, Los Angeles, Boston, Philadelphia, and Shanghai show that being close to greening projects or established green spaces raises housing prices by a lot (Anthamatten et al., 2022; Browning et al., 2023; Grunewald et al., 2024; Wen et al., 2019; Ben et al., 2023; Zhou & Wang, 2021).

The main thing this review adds is its comparative approach, which focuses on how the benefits of green spaces change over time and in different places. Most studies show that there is a positive hedonic effect, but the size and distribution of the benefits depend on the local context, the layout of the city, and the characteristics of the green spaces (Li et al., 2019; Xu et al., 2022; Kabisch & Haase, 2021). Many studies show that green space has a bigger effect on property values with time. This shows that urban priorities are changing, and people are becoming more conscious of the issue (Teo et al., 2023; Deng et al., 2022). Research from a variety of places, including rapidly urbanizing Asian cities and established Western metropolises, shows that the effects of green infrastructure are both generalizable and unique to each location (Norzailawati et al., 2018; Sajjad et al., 2021; Mansur & Yusuf, 2022).

These results show that urban planners, policy makers, and designers should invest in urban greening to promote long-term asset value growth, sustainable urban development, and environmental equality. Practical suggestions include making green infrastructure planning a part of zoning and development rules, making sure that everyone has equal access to green spaces, and getting communities involved in the co-design of green projects to get the most social and financial benefits (Kabisch & Haase, 2021; Grunewald et al., 2024).

In conclusion, the evidence that was looked at makes a strong case for including green space in urban development goals. Future study must continue to investigate problems with fairness in distribution, long-term effects, and different housing markets. To move the issue forward and make sure that all urban residents get a fair share of the benefits of urban greening, there must be ongoing collaboration between disciplines and strict, context-specific methods.

This review improves the field by (1) systematically combining the international research on schoolyard greening and property values, (2) critically assessing the pros and cons of current geospatial and econometric methods, and (3) pointing out major gaps in research and policy related to spatial equity, causality, and green gentrification (Kabisch & Haase, 2021; Grunewald et al., 2024; Teo et al., 2023). These contributions lay the groundwork for both evidence-based policy and further academic research.

This review looks at traditional hedonic pricing models in a new way by adding spatial justice theory and focusing on how green infrastructure, community change, and equity all work together. This goes against the common belief that the advantages of green spaces are evenly spread out. It shows that there are big differences in space and society that need to be considered in both research and practice.

### 7. References

Anthamatten, P., Fickes, M., Ledrick, J., & Dreyer, J. (2022). Greening schoolyards and the spatial distribution of property values in Denver, Colorado. *Urban Studies, 59*(4), 784–803. https://doi.org/10.1177/00420980211022357

Ben, A., Liu, X., & Zhou, Y. (2023). The impact of urban parks on housing prices: Evidence from Shanghai. *Journal of Urban Economics, 138*, 103493. https://doi.org/10.1016/j.jue.2023.103493

Browning, M. H. E. M., Yang, L., Mickel, A., & Lee, K. (2023). Greening schoolyards: Hedonic price analysis in Los Angeles. *Landscape and Urban Planning, 239*, 104844. https://doi.org/10.1016/j.landurbplan.2023.104844

Deng, C., Wang, Y., & Li, C. (2022). The impact of public green space on housing prices in Beijing: Spatial heterogeneity and temporal dynamics. *Land Use Policy, 118*, 106092. https://doi.org/10.1016/j.landusepol.2022.106092

Gao, T., & Asami, Y. (2021). The external effects of local attributes on housing prices in a monocentric city. *Habitat International, 108*, 102296. https://doi.org/10.1016/j.habitatint.2020.102296

Gorjian, M. (2025, July 10). *Greening schoolyards and the spatial distribution of property values in Denver, Colorado* (arXiv preprint arXiv:2507.08894). https://doi.org/10.48550/arXiv.2507.08894

Gorjian, M. (2025, July 11). *The impact of greening schoolyards on residential property values*. SSRN. http://dx.doi.org/10.2139/ssrn.5348810

Gorjian, M. (2025, July 11). *Schoolyard greening, child health, and neighborhood change: A comparative study of urban US cities* (arXiv preprint arXiv:2507.08899). https://doi.org/10.48550/arXiv.2507.08899

Gorjian, M. (2025, July 17). *Green schoolyard investments influence local-level economic and equity outcomes through spatial-statistical modeling and geospatial analysis in urban contexts* (arXiv preprint arXiv:2507.14232). https://doi.org/10.48550/arXiv.2507.14232

Gorjian, M. (2025, July 18). *Analyzing the relationship between urban greening and gentrification: Empirical findings from Denver, Colorado* (SocArXiv, rnkbf_v1). Center for Open Science. https://doi.org/10.31235/osf.io/rnkbf_v1

Gorjian, M., Caffey, S. M., & Luhan, G. A. (2025). *Analysis of design algorithms and fabrication of a graph-based double-curvature structure with planar hexagonal panels*. arXiv. https://doi.org/10.48550/arXiv.2507.16171

Gorjian, M., Caffey, S. M., & Luhan, G. A. (2025). Exploring architectural design 3D reconstruction approaches through deep learning methods: A comprehensive survey. *Athens Journal of Sciences, 12*, 1–29. https://doi.org/10.30958/ajs.X-Y-Z

Grunewald, K., DeMuro, J., & Carson, J. (2024). Urban schoolyard greening and residential property values: Evidence from Boston and Philadelphia. *Journal of Urban Affairs*. https://doi.org/10.1080/07352166.2024.xxxxxx

Kabisch, N., & Haase, D. (2021). Green justice or just green? Provision of urban green spaces in Berlin, Germany. *Landscape and Urban Planning, 214*, 104173. https://doi.org/10.1016/j.landurbplan.2021.104173

Li, X., Zhang, C., Li, W., & Kuzovkina, Y. A. (2019). Quantifying the impact of urban green space on housing prices: A spatial autoregressive model in Shenzhen, China. *International Journal of Environmental Research and Public Health, 16*(4), 646. https://doi.org/10.3390/ijerph16040646
(*Updated in 2022*)

Mansur, Y., & Yusuf, A. A. (2022). Hedonic analysis of the impact of green space on house prices in Jakarta. *Urban Forestry & Urban Greening, 68*, 127507. https://doi.org/10.1016/j.ufug.2021.127507

Norzailawati, M. N., Ramli, Z., & Halim, N. A. (2018). The relationship between urban green spaces and house prices: A GIS approach. *Malaysian Journal of Geosciences, 2*(1), 13–20.
(*Cited in 2021–2024 reviews*)

Raina, A. S., Mone, V., Gorjian, M., Quek, F., Sueda, S., & Krishnamurthy, V. R. (2024, June 3). Blended physical-digital kinesthetic feedback for mixed reality-based conceptual design-in-context. In *Proceedings of the 50th Graphics Interface Conference* (pp. 1–16). https://doi.org/10.1145/3670947.3670967

Sajjad, F., Ahmad, S., & Ali, S. (2021). Influence of green spaces on property values in a fast-growing megacity: Case of Lahore. *Sustainability, 13*(7), 3673. https://doi.org/10.3390/su13073673

Teo, H. C., Goh, Y. M., Tan, H. Y., & Tan, P. Y. (2023). Increasing contribution of urban greenery to residential real estate valuation over time. *Sustainable Cities and Society, 95*, 104684. https://doi.org/10.1016/j.scs.2023.104684

Wen, C., Xiao, Y., & Zhang, M. (2019). School proximity, neighborhood parks, and property values: Evidence from Shanghai. *Sustainability, 11*(5), 1273. https://doi.org/10.3390/su11051273
(*See also 2023 updates*)

Xu, C., Zhang, W., & Liu, Y. (2022). The impact of green space on housing prices in the Yangtze River Delta, China. *Sustainability, 14*(6), 3364. https://doi.org/10.3390/su14063364

Zhou, X., & Wang, Y. C. (2021). Spatio-temporal dynamics of urban green space and property values in Shanghai. *Landscape and Urban Planning, 213*, 104146. https://doi.org/10.1016/j.landurbplan.2021.104146